\magnification=1200
\parskip 10pt plus 5pt
\parindent 14pt
\baselineskip=18pt
\input mssymb
\pageno=0
\footline{ \ifnum \pageno<1 \else \hss \folio \hss \fi }
\line{\hfil{DAMTP-R/95/29~~~~~~}}
\line{\hfil{May, 1995~~~~~~~~~~~~~~~~}}
\vskip 1in
\centerline{\bf INFINITE-GENUS SURFACES AND THE UNIVERSAL GRASSMANNIAN}
\vskip .75in
\centerline{\bf SIMON DAVIS}
\vskip .5in
\centerline{Department of Applied Mathematics and Theoretical Physics}
\vskip 5pt
\centerline{University of Cambridge}
\vskip 5pt
\centerline{Silver Street, Cambridge CB3 9EW}
\vskip .65in
\noindent{\bf ABSTRACT.} Correlation functions can be calculated on Riemann
surfaces using the operator formalism.  The state in the Hilbert space of the
free field theory on the punctured disc, corresponding to the Riemann surface,
is constructed at infinite genus, verifying the inclusion of these
surfaces
in the Grassmannian.  In particular, a subset of the class of $O_{HD}$
surfaces can be identified with a subset of the Grassmannian.  The
concept of
flux through
the ideal boundary is used to study the connection between infinite-genus
surfaces and the domain of string perturbation theory.  The different roles of
effectively closed surfaces and surfaces with Dirichlet boundaries in a
more complete formulation of string theory are identified.

\vfill
\eject

It is known that higher-genus correlation functions for scalar or
free-fermion fields
can be constructed in the operator formalism [1][2].  In this formulation, for
every Riemann surface of finite genus with boundary, one may associate a state
$\vert \phi \rangle$ in the Hilbert space based on the boundary by performing
the path integral of the conformal field theory with the values of the
fields specified on the boundary.  The state $\vert \phi \rangle$ encodes
information about the Riemann surfaces and one can obtain correlation
functions by inserting vertex operators between $\langle 0 \vert$, the standard
vacuum in the Hilbert space, and $\vert \phi \rangle$.  While this actually
only allows the computation of the correlation functions when the positions
of the vertex operators are on the disk, it is possible to
analytically
continue
the expressions to the whole Riemann surface.

In this paper, the operator formalism is generalized for the first time to
infinite-genus surfaces.  These surfaces were originally discussed in the
context of string theory in [3], where the total partition function is
expressed
as the section of a bundle on universal moduli space.  While the inclusion of
infinite-genus surfaces in universal moduli space suggests the presence of
non-perturbative effects, it is necessary to make the formulation precise, by
distinguishing between the domain of string perturbation theory and the
non-perturbative contribution.  This has been a central problem in string
theory since the initial analyses of perturbative scattering amplitudes.
It can be shown that there is a class of infinite-genus surfaces which may be
included in the perturbative expansion of the vacuum amplitude known as the
set of $O_G$ or effectively closed surfaces [4][5][6].

It therefore follows that the introduction of infinite-genus surfaces does not
lead to non-perturbative effects, a result which had not yet been established
at the time that the article [3] appeared.  With an appropriate domain for
string perturbation theory, including effectively closed infinite-genus
surfaces, the perturbative expansion of the S-matrix can be defined by the
vertex operator and moduli space integrals associated with these surfaces.
The development of a complete formulation of string theory motivates the
search for the source of non-perturbative effects.  An elegant approach,
recently advanced in [7][8]
\hfil\break
[9][10][11], involves the association of
non-perturbative effects with the insertion of Dirichlet boundaries in the
string worldsheet.

The connection between infinite-genus surfaces and the domain of string
perturbation theory is analyzed using the concept of flux through the ideal
boundary and divergences in the amplitudes as vertex operators
approach the boundaries.
These properties can be used to distinguish between the effectively closed
surfaces and the surfaces with Dirichlet boundaries.
For the effectively closed surfaces, it will be demonstrated that
a vanishing flux condition is satisfied,
which allows for the states $\vert \phi \rangle$ to be solved exactly and
the surfaces to be included in the universal Grassmannian in the same manner
as closed finite-genus surfaces, implying that these surfaces may be used in
the perturbative expansion of the S-matrix.
Support for the
association of non-perturbative effects with surfaces having Dirichlet
boundaries shall come from the behaviour of the correlation function as the
vertex operators approach the boundaries.  The different roles of the
effectively closed surfaces and the surfaces with Dirichlet
boundaries, therefore, can be established in a more complete
formulation of string theory.

In [1], it was shown that for the free scalar field theory on a
Riemann surface,
$$S~=~\int_{\Sigma-D}~X \partial {\bar \partial} X
\eqno(1)$$
there are an infinite number of symmetries
$$X \to X~+~\epsilon_n X_n
\eqno(2)$$
where $X_n$ represents a perturbation which corresponds to a function
with a
pole of order n at $t = 0$ that is harmonic in
$\Sigma~-~D$ (Fig. 1).

\vskip 0.2in
\input epsf.tex

\vbox{
\epsfysize=2.4in
\centerline{\epsfbox{pole.eps}}
\vskip 0.2in
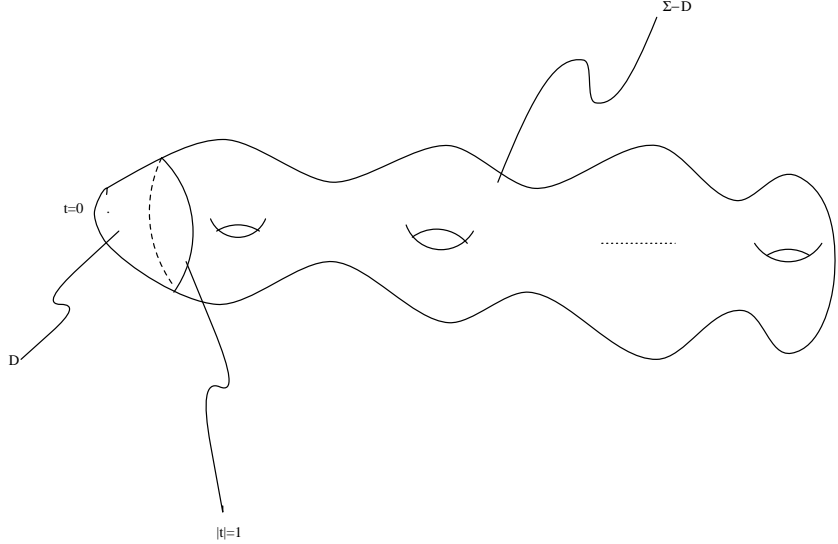
\noindent{\bf Fig. 1.  Splitting of a Riemann surface into a disc D and the
genus-g complement $\Sigma - D$.}}

The action becomes
$$\int_{\Sigma - D}~X \partial {\bar{\partial}} X~+~\epsilon_n
\int_{\Sigma-D} \partial (X_n^a {\bar \partial}
X )~+~\epsilon_n \int_{\Sigma-D} {\bar \partial} (X_n^h \partial X)
{}~=~\int_{\Sigma-D} X \partial {\bar \partial} X~+~\epsilon_n Q_n
\eqno(3)$$
where
$$Q_n~=~\int_S~(X_n^a {\bar \partial}X~+~X_n^h \partial X)
\eqno(4)$$
with $X_n^h$ and $X_n^a$ being the holomorphic and anti-holomorphic parts of
$X_n$ respectively.  It is only when the scalar field X satisfies the
equations of motion and is therefore harmonic that the decomposition
$X~=~X^h(z)~+~X^a({\bar z})$ is valid, simplifying the formula for the charge
$Q_n$.
The path integral does not change if
$$Q_n \vert \phi \rangle~=~0
\eqno(5)$$
This leads to an infinite number of conditions which are integrable because
the charges $Q_n$ commute.  The state $\vert \phi \rangle$ can be obtained
within a constant factor.

This argument made use of Stokes' theorem.  Now consider the planar covering
of a finite-genus surface with boundary S (Fig. 2)

\vskip 0.2in
\input epsf.tex

\vbox{
\epsfysize=2.2in
\centerline{\epsfbox{Schottky.eps}}
\vskip 0.2in
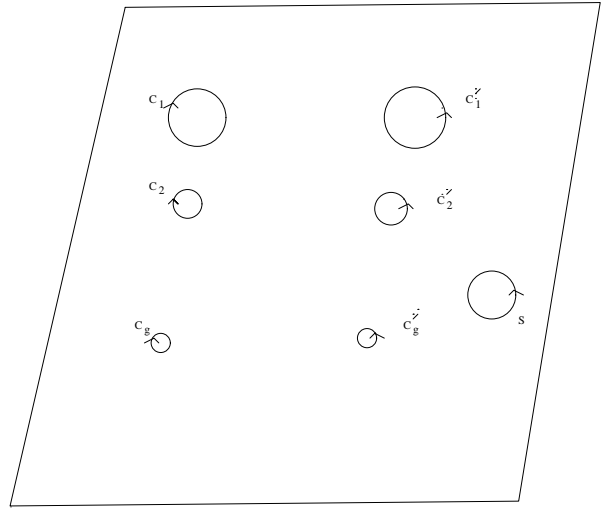
\noindent{\bf Fig. 2 Cancellation of contour integration on the
Schottky
covering of a Riemann surface.}}

\noindent One sees that the boundary of the fundamental region is
$S\cup \cup_{n=1}^g
{}~(C_n \cup C_n^\prime)$, the union of two distinct sets, S and $\cup_{n=1}^g
(C_n \cup C_n^\prime)$, where the latter set represents the 2g isometric
circles of the Schottky group uniformizing the surface.  The
differential on
the Riemann surface is now
replaced by a differential on the plane which is automorphic under the
action of the Schottky group, $\omega (\Gamma z)~=~\omega(z)$.
However, the contours $C_n$ and $C_n^\prime$ are oriented in opposite
directions.  Consequently, the line integrals over them vanish and one is left
with a line integral over S.

There also exist points in the universal Grassmannian, defined to be
the set
of rays in the Hilbert
space of states for the free field theory on the punctured disc,
corresponding to infinite-genus surfaces, and one may wish to solve for
the state $\vert \phi \rangle$ using the method applied to finite-genus
surfaces.  The use of Stokes' theorem implies that one would have to consider
the line integral on S and the integral on the ideal boundary (Fig. 3).

\vskip 0.2in
\input epsf.tex
\vbox{
\epsfxsize= 0.8 \hsize
\centerline{\epsfbox{boundary.eps}}
\vskip 0.2in
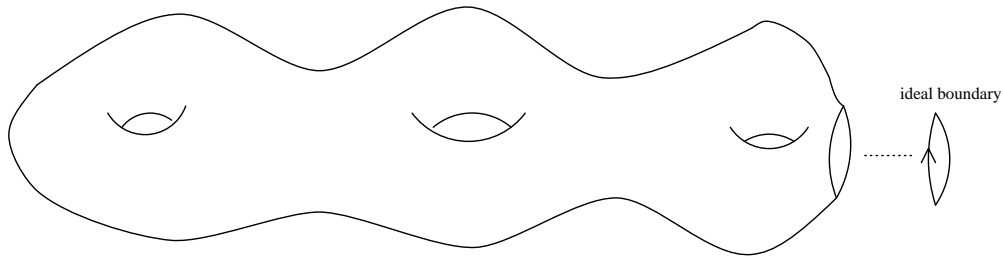
\noindent{\bf Fig. 3 Contour integration over the ideal boundary of a Riemann
surface.}}

For integrability, it is necessary to impose the condition
$$\int_{{ideal}\atop {bdy}}~j_n~=~0
\eqno(6)$$
where $j_n~=~X_n^a {\bar \partial} X~+~X_n^h \partial X$ is the current.
The vanishing of the integral at the ideal boundary is needed to satisfy
the operator condition $Q_n \vert \phi \rangle~=~0$ (5) as $Q_n~=~
\int~j_n$.
  The simplest way to achieve this result follows
by requiring that $X_n$ vanishes at the boundary for each n.

Recalling that for compact surfaces, by the Weierstrass gap theorem, there
are g values of n between 1 and 2g for which there does not exist a
meromorphic function with a pole of order n at some point.  This problem is
circumvented by allowing $X_n$ to have holomorphic and anti-holomorphic
parts [1].

The Weierstrass gap problem does not occur for infinite-genus surfaces.
Every divisor on a non-compact surface is the divisor of a meromorphic
function.  In fact, given an arbitrary discrete set of points on the surface,
one can find a meromorphic function which has poles of any order that is
selected for these points.  This follows from the Behnke-Stein theorem [12]
which shows that there exists an everywhere regular analytic function
with any discrete set of zeros.  The order of these zeros can be chosen
arbitrarily.  If $f$ is such an analytic function, then ${1\over f}$ is
a meromorphic function with the required properties.  Moreover, it is
possible that $f$ is unbounded in the approach to the ideal boundary.
Then ${1\over f}$ would tend to zero.  So, it may be possible to find
meromorphic functions $X_n$, with poles of order  n, tending to zero at the
ideal boundary, for each n.  This would imply that the state
$\vert \phi \rangle$ would be a simple tensor product of a state with
holomorphic degrees of freedom and a state with anti-holomorphic degrees
of freedom.

Thus, defining the universal Grassmannian to be the space of such
vacuum states,
up to a multiplicative factor, every compact finite-genus surface and
certain infinite-genus surfaces will correspond to points in the Grassmannian.
Moreover, the one-to-one correspondence between the standard Fock vacuum
and the vacuum state for the Riemann surface could be regarded as
being
equivalent to the formulation of the axiom of asymptotic completeness
in
Euclidean quantum gravity, which is required for the factorization of
the
${\$}$ matrix [13].  This one-to-one correspondence between the vacua
implies
that
unitarity of the free fermion theory is maintained in the
infinite-genus
limit.   The extension to string theory might be established by
consideration
of the
full string action.

An example of an infinite-genus surface which could be included in the
universal
Grassmannian would be the sphere with an infinite number of handles studied
in [5].  One may recall that the correlation function can be
calculated
explicitly
using the method of images.
$$\eqalign{G(z_P;z_R, z_S)~&=~\sum_\alpha~ln \left \vert {{z_P~-~V_\alpha z_R}
\over {z_P~-~V_\alpha z_S}} \right\vert
\cr
{}~&~~~~~~~-~{1\over {2 \pi}}
{}~\sum_{m,n=1}^\infty~Re(v_n(z_P))~(Im~\tau)^{-1}_{mn}~Re(v_n(z_R)~-~v_n(z_S))
\cr}
\eqno(7)$$
where
$$v_n(z)~=~\sum_\alpha~^{(n)}~ln\left({{z~-~V_\alpha \xi_{1n}}
\over {z~-~V_\alpha
 \xi_{2n}}} \right)
\eqno(8)$$
is holomorphic on the the fundamental domain since $\xi_{1n},~\xi_{2n}$
are fixed points of $T_n$ and $\sum_\alpha~^{(n)}$ includes only those
$V_\alpha$ for which $T_n,~T_n^{-1}$ is not the right-most member of the
element $V_\alpha$.
It may be noted that in the second term, the definition of $v_n(z)$ requires
convergence of the Poincare series [5].  While the period matrix is
infinite-dimensional, it still has eigenvalues
$$(Im~\tau)~{\tilde e}_n~=~\lambda_n~{\tilde e}_n
\eqno(9)$$
where ${\tilde e}_n$ is the nth eigenvector, which can also be written as
${\tilde e}_n~=~U~e_n$ with $e_n$ being the nth unit vector in
${\Bbb R}^\infty$.  As
$U^{-1} (Im~\tau) U~=~diag(\lambda_1,...,\lambda_n,...)$,
\hfil\break
$(Im~\tau)^{-1}~=~U~diag(\lambda_1^{-1},...,\lambda_n^{-1},...) U^{-1}$
if the eigenvalues are non-zero.  The existence of $U^{-1}$ requires that
the column vectors of U, eigenvectors of $(Im~\tau)$, be linearly independent.
This follows since the complement to the space spanned by the eigenvectors is
the kernel, which will have non-zero dimension if the eigenvectors are
linearly dependent.  However, if $dim~ker(Im~\tau) \ne 0$, the determinant of
$Im~\tau$ vanishes, contrary to the assumption that the eigenvalues are
non-zero.

Positivity of the eigenvalues of $Im~\tau$ generally follows from the
bilinear relation
$$(\omega,~\sigma)~=~\sum_{n=1}^g~\left[~\int_{A_n}~\omega~\int_{B_n}~
{\bar \sigma}~-~\int_{A_n}~\sigma~\int_{B_n}~{\bar \omega}~\right]
\eqno(10)$$
A generalized bilinear relation holds on infinite-genus surfaces in the class
$O_{HD}$ [14], which, by definition, admit no non-constant harmonic functions
with finite
Dirichlet norm.  To verify this property for open surfaces can be quite
involved, and it is easier to use the following result [15]:

\noindent ${\underline{Theorem}}$. Let $p_0,~p_1$ be principal
functions for any pairs of logarithmic singularities
\hfil\break
$log \vert z~-~z_R
\vert$ and $-log \vert z~-~z_S\vert$.  If $p_0~-~p_1$ is constant, then the
surface is in $O_{HD}$.

The principal functions are harmonic everywhere except for the singularities,
with $p_0$ having zero normal derivative and $p_1$ having vanishing flux
on the ideal boundary.  This theorem will now be used for surfaces which can
be uniformized by groups of Schottky type.

To the Green function (7), one may add a harmonic
function of the form
\hfil\break
$\sum_n~a_n~\sum_\alpha~Re~v_n(V_\alpha z)~+~C$.
The normal derivative at a point on  a circle of radius r is
$$\eqalign{{1\over 2}~\sum_n~a_n&~\sum_\alpha~\left(e^{i\theta}
{{d v_n(u)}\over {du}}\bigg\vert_{u=V_\alpha z}~+~c.c. \right)
\cr
&~=~{1\over 2}~\sum_n~a_n~\sum_\alpha~\left( e^{i \theta}~
\sum_\beta~^{(n)}~{{V_\beta \xi_{1n}~-~V_\beta \xi_{2n}}
\over {(V_\alpha z~-~V_\beta \xi_{1n})
(V_\alpha z~-~V_\alpha \xi_{2n})}}~+~c.c.~\right)
\cr}
\eqno(11)$$
If $\infty$ is a limit point of the uniformizing Schottky group $\Gamma$,
invariance of the limit point set $\Gamma$ implies that there exist elements
$V_\alpha$
such that $V_\alpha z~=~V_\beta \xi_{1n}$ or $V_\alpha z~=~ V_\beta \xi_{2n}$.
These two terms are infinite but they cancel.  Thus, the terms giving the
dominant contribution are those for which $V_\alpha z \in I_{V_\beta^{-1}}$,
requiring $V_\alpha~=~V_\beta~V_\gamma^\prime$ where the left-most members
of $V_\gamma^\prime$ do not form the element $V_\beta^{-1}$.
$$\eqalign{\sum_\gamma~^{(\beta)}~\sum_\beta~^{(n)}~
{{V_\beta \xi_{1n}~-~V_\beta \xi_{2n}}\over {(V_\beta V_\gamma^\prime z
{}~-~V_\beta \xi_{1n}) (V_\beta V_\gamma^\prime z~-~V_\beta \xi_{2n})}}
{}~&=~
\cr
\sum_\gamma~^{(\beta)}~\sum_\beta~^{(n)}&~~{{(\xi_{1n}-\xi_{2n})
\gamma_\beta^2}\over {{{(V_\gamma^\prime z~-~\xi_{1n})}\over
{\left(V_\gamma^\prime z~+~{{\delta_\beta}\over
{\gamma_\beta}}\right)}}
{{(V_\gamma^\prime z~-~
\xi_{2n})}\over {\left(V_\gamma^\prime z~+~{{\delta_\beta}
\over {\gamma_\beta}}\right)}} } }
\cr}
\eqno(12)$$

As z tends to $\infty$, $V_\gamma^\prime z~\equiv~{{\alpha_\gamma z~+~
\beta_\gamma}
\over {\gamma_\gamma z~+~\delta_\gamma}}$ tends to ${{\alpha_\gamma}\over
 {\gamma_\gamma}}$ so that the terms in (11) become constant.  Since the
circle at infinity represents the ideal boundary, defined to be the limit set
of $\Gamma$ factored by $\Gamma$, the normal derivative vanishes at the
boundary only if each $a_n$ is zero.  The flux is formally obtained by
integrating the normal derivative around the circle at infinity, which
could
be viewed as
potentially representing the ideal boundary of a non-compact surface.
Expanding the sum in (12) in
powers of ${1\over z}$, $c_{0n}~+~{{c_{1n}}\over z}~+~{{c_{2n}}\over {z^2}}
{}~+~...$ and using $\int_0^{2\pi}~r~e^{i\theta} c_0 d \theta~=~0$,
$lim_{r \to \infty}~\int_0^{2\pi}~r e^{i \theta} {{c_{mn}}\over {z^m}}~d \theta
{}~=~0$
for $m~\ge~2$, one sees that the contribution of (12) to the flux is
$2 \pi c_{1n}$ where
$$\eqalign{c_{1n}~=~\sum_\gamma~^{(\beta)}&~\sum_\beta~^{(n)}~
(\xi_{1n}-\xi_{2n})~ \gamma_\beta^2~ \left({{\alpha_\gamma}\over
{\gamma_\gamma}}~+~{{\delta_\beta}\over {\gamma_\beta}}\right)
{}~{{\beta_\gamma~-~\delta_\gamma}\over {\gamma_\gamma}}~
{1\over {\left({{\alpha_\gamma}\over {\gamma_\gamma}}~-~\xi_{1n}\right)}}~
\cr
&{1\over {\left({{\alpha_\gamma}\over {\gamma_\gamma}}~-~\xi_{2n}\right)}}
\left[~2~-~\left({{\alpha_\gamma}\over {\gamma_\gamma}}~+~{{\delta_\beta}\over
{\gamma_\beta}}\right)
\left[{1\over {\left({{\alpha_\gamma}\over
{\gamma_\gamma}}-\xi_{1n}\right)}}
{}~+~
{1\over {\left({{\alpha_\gamma}\over {\gamma_\gamma}}~-~\xi_{2n} \right)}}
\right]\right]
\cr}
\eqno(13)$$
Thus, again, to obtain vanishing flux, one requires that all $a_n$ vanish.
It follows that the difference between the principal functions $p_0~-~p_1$
is a constant and the surface is in the class $O_{HD}$.

Since $v_n(z)$ is an automorphic function on the complex plane, $\sum_\alpha~
Re~v_n(V_\alpha z)$ diverges and should only be regarded as a formal
expression.
Alternatively, one may consider the functions
$$h_n(z, {\bar z})~=~\sum_\alpha \sum_p~Re[v_n(V_\alpha z)~-~v_n(V_\alpha z_0)]
{}~b_p~(V_\alpha z~-~V_\alpha z_0)^p
\eqno(14)$$
where each term is finite for $p\ge 1$ because of the convergence of the
Poincare series $\sum_{\alpha\ne I} \vert \gamma_\alpha\vert^{-2}~<~\infty$.
Thus,
$$\eqalign{h_n^\prime(z, {\bar z})~&=~\sum_\alpha \sum_p~{d\over {du}}
\left[[Re(v_n(u))~-~
Re(v_n(V_\alpha(z_0)))]~b_p~(u~-~V_\alpha z_0)^p \right]
\bigg\vert_{u=V_\alpha z}
\cr
{}~&=~\sum_\alpha \sum_p\sum_\beta~^{(n)}~Re~\left[{{V_\beta \xi_{1n}~-~
V_\beta \xi_{2n}}\over {(V_\alpha z~-~V_\alpha \xi_{1n}) (V_\alpha z~-~
V_\alpha \xi_{2n})}} \right]~b_p (V_\alpha z~-~V_\alpha z_0)^p
\cr
&~+~\sum_\alpha \sum_p~[Re(v_n(V_\alpha z)~-~v_n(V_\alpha z_0))]~p~b_p~
(V_\alpha z~-~V_\alpha z_0)^{p-1}
\cr}
\eqno(15)$$
and
$$\eqalign{h_n^\prime(z, {\bar z})~{\rightarrow\atop {z\to \infty}}~&
\sum_p~b_p~(z-z_0)^p~\sum_\beta~^{(n)}~Re~\left[{{V_\beta
\xi_{1n}~-~V_\beta
\xi_{2n}}
\over {(z~-~V_\beta \xi_{1n})(z~-~V_\beta \xi_{2n})}}\right]
\cr
&~+~\sum_p~p~b_p~(z-z_0)^{p-1}~[Re(v_n(z)~-~v_n(z_0))]
\cr}
\eqno(16)$$
implying divergence of the integral
$$\int\int_{F.D.}~\vert h_n^\prime(z, {\bar z})\vert^2~dz~\wedge d{\bar z}
{}~\to~\sum_{r=0}^\infty~{\tilde b}_r~(Re(v_n(z_0))^2~\int\int_{F.D.-\Delta}~
\vert z~-~z_0\vert^{r-2}~dz~\wedge d{\bar z}~+~finite
\eqno(17)$$
with F.D. being the fundamental domain of the uniformizing group,
$\Delta$ being a region of finite size and ${\tilde
b}_r~=~\sum_{p+q=r}~p~q~
b_p~b_q$.  If this property
is valid for all non-constant harmonic functions on the Riemann surface, then
it belongs to the class $O_{HD}$.  The category of Riemann surface can also be
determined by the normal derivative and the flux for the harmonic functions
$h_n(z,{\bar z})$ of equation (14).  By analogy with equation (12), the
normal derivative in the limit $z \to \infty$ is
$$\eqalign{&{1\over 2}~\sum_n~a_n~e^{i\theta}~h_n^\prime(z)~+c.c.
\cr
& ={1\over 2}\sum_na_ne^{i\theta}\sum_p\sum_\gamma~^{(\beta)}~\sum_\beta~^{(n)}
{}~{{(\xi_{1n}-\xi_{2n})~\gamma_\beta^{2-2p}~b_p~\left({{\alpha_\gamma}\over
{\gamma_\gamma}}~-~V_\gamma^\prime z_0\right)^p}
\over {\left({{\alpha_\gamma}\over {\gamma_\gamma}}-\xi_{1n}\right)
\left({{\alpha_\gamma}\over {\gamma_\gamma}}-\xi_{2n}\right)
\left({{\alpha_\gamma}\over {\gamma_\gamma}}+{{\delta_\beta}\over
{\gamma_\beta}}\right)^{p-2}\left(V_\gamma^\prime z_0 + {{\delta_\beta}
\over {\gamma_\beta}}\right)^p}}
\cr
& ~+~{1\over 2}\sum_n~a_n~e^{i\theta}~\sum_p~\sum_\gamma~^{(\beta)}~
\sum_\beta~^{(n)}~\sum_\delta~^{(n)}~ln\left\vert {{
{{\alpha_\gamma}\over
{\gamma_\gamma}}
{}~-~V_\delta \xi_{1n}}\over { {{\alpha_\gamma}\over {\gamma_\gamma}}
{}~-~V_\delta \xi_{2n}  }}~{{ V_\gamma^\prime z_0~-~
V_\delta \xi_{2n}}\over {V_\gamma^\prime~z_0~-~V_\delta \xi_{1n}} }
\right\vert
\cr
&~~~~~~~~~~~~~~~~~~~~~~~~~~~~~~~~~~~~~~~~~~~~~~\cdot
p~b_p~\left[{{\left(
{{\alpha_\gamma}\over {\gamma_\gamma}}-
V_\gamma^\prime~z_0\right)}
\over {\left({{\alpha_\gamma}\over {\gamma_\gamma}}+{{\delta_\beta}\over
{\gamma_\beta}}\right)\left(V_\gamma^\prime~z_0+{{\delta_\beta}\over
{\gamma_\beta}}\right)}}\right]^{p-1}\gamma_\beta^{2-2p}
+c.c.
\cr}
\eqno(18)$$
As each of the terms multiplying the coefficients $a_n$ are non-vanishing
constants, the normal derivative is zero only if each of the $a_n$ is
zero.  Note that the sums multiplying $a_n$ will only be finite when
$b_0~=~b_1~=~0$ and $\sum_\beta~^{(n)}~\gamma_\beta^{2-2p}~<~\infty$
only when $p \ge 2$.

The flux is $2 \pi~\sum_n~a_n c_{1n}$ where, by analogy with equation (13),
$$\eqalign{c_{1n}~=~\sum_\gamma~^{(\beta)}~\sum_\beta~^{(n)}&~
{{\gamma_\beta^{-2p}}\over {\left(z_0~+~{{\delta_\beta}\over {\gamma_\beta}}
\right)^p}}~\left({{\alpha_\gamma}\over {\gamma_\gamma}}~+~{{\delta_\beta}
\over {\gamma_\beta}}\right)^{-(p-1)}~\left(V_\gamma^\prime z_0~+~
{{\delta_\beta}\over {\gamma_\beta}}\right)^{-p}~
\cr
&\cdot{1\over {\left({{\alpha_\gamma}\over {\gamma_\gamma}}-\xi_{1n}\right)}}
{}~{1\over {\left({{\alpha_\gamma}\over {\gamma_\gamma}}-\xi_{2n}\right)}}
\cr
&\Bigg[{{\beta_\gamma - \delta_\gamma}\over {\gamma_\gamma}}
\left[2-\left({{\alpha_\gamma}\over {\gamma_\gamma}}+{{\delta_\beta}
\over {\gamma_\beta}}\right)\left[{1\over {\left({{\alpha_\gamma}\over
{\gamma_\gamma}}-\xi_{1n}\right)}}+{1\over {\left({{\alpha_\gamma}\over
{\gamma_\gamma}}-\xi_{2n} \right)}} \right]\right]
\cr
&~-~p\left[ \left(z_0+{{\delta_\beta}
\over {\gamma_\beta}}\right)\left({{\alpha_\gamma}\over {\gamma_\gamma}}
+{{\delta_\beta}\over {\gamma_\beta}}\right)
+{{\beta_\gamma-\gamma_\gamma}\over {\delta_\gamma}}\right]\Bigg]
\cr}
\eqno(19)$$
Actually, by Green's theorem, the flux for a harmonic function equals
the
Dirichlet norm, and therefore,
the sums in (19) should diverge.
The flux vanishes only when $a_n~=~0$ for all n, and consequently the
difference
between the principal functions $p_0$ and $p_1$ must be a constant, implying
that
the surface is in $O_{HD}$.

Indeed, it has been proven that surfaces uniformized by a group of
Schottky
type
are in the class $O_G$ [4], characterized by the property that the ideal
 boundary
has zero linear measure.  The model considered here is useful for consideration
of both $O_G$ surfaces and the surfaces with ideal boundary at infinity, as the
isometric circles are clearly tending towards a point $\infty$ in the
extended complex plane, but this
may also be regarded as the infinite radius limit of a sequence of
circles,
with each circle representing a closed
curve on the Riemann surface.  The analysis above, derived for surfaces
uniformized  by groups of Schottky type, and therefore applicable to $O_G$
surfaces, provides a useful indication of the calculation of the
flux for the more general $O_{HD}$ surfaces.  The restriction to the $O_G$
surfaces can be implemented by taking the radius of the limit circle
to be
zero,
and one may notice that this factor is cancelled in the calculation of the
flux so that its magnitude does not affect the results mentioned above.

The same method can be applied to the Green function (7) for $O_G$ surfaces
uniformized by Schottky groups.  One may also recall that the
existence of a Green function with two sources at $z_R$ and $z_S$, with
behaviour $ln \vert z_P-z_R\vert$ and $ -ln \vert z_P - z_S \vert$ can be
demonstrated for more general Riemann surfaces [16].  In the limit
$z_P \to \infty$,
$G(z_P;z_R,z_S)$ and ${{\partial G}\over {\partial z_P}}$ tend to zero.
Similarly, one may wish to calculate the flux at the accumulation point of
the isometric circles from this Green
function.  The study of the principal functions shows that this
quantity also
vanishes.
It is now possible to make an identification of a
subset of $O_{HD}$ with a subset of the Grassmannian.
A necessary condition for the solvability of equation (5) on open surfaces
with ideal boundary is the vanishing of the flux (6).  Those surfaces
in $O_{HD}$
for which this condition is sufficient will belong to the Grassmannian because
an exact solution for $\vert \phi \rangle$ has been obtained.
Conversely, those surfaces corresponding to points in the Grassmannian
satisfy the vanishing flux condition and therefore possess Green functions
with two sources with vanishing normal derivative at the ideal boundary
if they also lie in $O_{HD}$.

One may note a similarity with the analysis of Green and Polchinski [9][10][17]
regarding the insertion of boundaries in string worldsheets.  It is of interest
to note that a divergence in the correlation function arises, for the surfaces
that they study, as the locations of the vertex operators near
the boundary [8].  This divergence confirms the interpretation of the
boundary
states as point-particle states [7][8].  The absence of similar
divergences in the computation of the four-point function [5] on a sphere with
an
infinite
number of handles is consistent with the absence of flux through the
accumulation point and the assumption that there are no additional sources
which may be associated with boundary states.  Consequently, the effectively
closed surfaces naturally may be included in the perturbative
expansion of
the S-matrix,
whereas the surfaces with Dirichlet boundaries inserted contribute to
non-perturbative effects in string theory.
\vskip .5in

\centerline{\bf Acknowledgements}
The vanishing flux condition at the ideal boundary, required for the
application
of the operator formalism to infinite-genus surfaces, was initially formulated
at the Lyman Laboratory in 1988.  I would like to thank
Prof. S. W. Hawking and Dr. G. W. Gibbons for their support while this paper
has been completed.

\vfill\eject

\centerline{\bf REFERENCES}

\item{[1]} C. Vafa, Phys. Lett. ${\underline{190B}}$ (1987) 47 - 54
\item{[2]} L. Alvarez-Gaume, C. Gomez and C. Reina, Phys. Lett.
${\underline{190B}}$ (1987) 55 - 62
\hfil\break
L. Alvarez-Gaume, C. Gomez, G. Moore and C. Vafa, Nucl. Phys.
${\underline{B303}}$ (1988) 455 - 521
\item{[3]} D. Friedan and S. Shenker, Phys. Lett. ${\underline {B175}}$
(1986) 287 - 296
\hfil\break
D. Friedan, ${\underline{Physics~and~Mathematics~of~Strings:}}$
\hfil\break
${\underline{Memorial~Volume~
{}~for~Vadim~Knizhnik}}$, ed. by L. Brink, D. Friedan and
\hfil\break
A. M. Polyakov (Singapore: World Scientific, 1990)
\item{[4]} S. Davis, Class. Quantum Grav. ${\underline{6}}$ (1989) 1791 - 1803
\item{[5]} S. Davis, Mod. Phys. Lett. A ${\underline {9}}$(14) (1994)
1299 - 1307
\item{[6]} S. Davis, J. Math. Phys. ${\underline{36}}$(2) (1995) 648 - 663
\item{[7]} M. B. Green, Phys. Lett. ${\underline{B266}}$ (1991) 325 - 336
\item{[8]} M. B. Green, Phys. Lett. ${\underline{B329}}$ (1994) 435 - 443
\item{[9]} M. B. Green and J. Polchinski, Phys. Lett. ${\underline{B335}}$
(1994) 377 - 382
\item{[10]} J. Polchinski, Phys. Rev. ${\underline{D50}}$ (1994) 6041 - 6045
\item{[11]} M. B. Green, `A Gas of D-Instantons', hep-th/9504108
\item{[12]} O. Forster, ${\underline{Lectures~on~Riemann~Surfaces}}$
(New York: Springer-Verlag, 1981)
\item{[13]} S. W. Hawking, Commun. Math. Phys. ${\underline{87}}$ (1982)
395 - 415
\item{[14]} R. D. M. Accola, Trans. Amer. Math. Soc., ${\underline{96}}$
(1960) 143
\hfil\break
Y. Kusonoki, Mem. Coll. Sci. Univ. Kyoto, Series A Math. ${\underline{30}}$(1)
1
\item{[15]} L. Ahlfors and L. Sario, ${\underline{Riemann~Surfaces}}$
(Princeton: Princeton University Press, 1960)
\item{[16]} H. Farkas and I. Kra, ${\underline{Riemann~Surfaces}}$
(New York: Springer-Verlag, 1980)
\item{[17]} M. B. Green and P. Wai, Nucl. Phys. ${\underline{B431}}$
(1994) 131 - 172

\end